\documentclass[pdflatex,sn-basic,onecolumn]{sn-jnl}

\usepackage{graphicx}%
\usepackage{multirow}%
\usepackage{amsmath,amssymb,amsfonts}%
\usepackage{amsthm}%
\usepackage{mathrsfs}%
\usepackage[title]{appendix}%
\usepackage{xcolor}%
\usepackage{textcomp}%
\usepackage{manyfoot}%
\usepackage{booktabs}%
\usepackage{algorithm}%
\usepackage{algorithmicx}%
\usepackage{algpseudocode}%
\usepackage{listings}%

\theoremstyle{thmstyleone}%
\theoremstyle{thmstyletwo}%

\theoremstyle{thmstylethree}%

\newcommand{\ion}[2]{\mbox{#1\,{\sc #2}}}%
\newcommand{\msun}{\mbox{$M_\odot$}}%

\begin{document}

\title{\Large How Should We Understand Core Mass Function?}

\subtitle{\large A Memo of the CMF2IMF Conference at ESO Garching}


\author*[1]{\fnm{Fengwei} \sur{Xu}} 

\author[2]{\fnm{Roberto} \sur{Galv{\'a}n-Madrid}} 

\author[3]{\fnm{Kaho} \sur{Morii}} 

\author[4]{\fnm{Thomas} \sur{Nony}} 

\author[2]{\fnm{Aina} \sur{Palau}}

\author[5]{\fnm{Alessio} \sur{Traficante}} 

\author[5,6]{\fnm{Alice} \sur{Nucara}}

\affil*[1]{Max Planck Institute for Astronomy, K{\"o}nigstuhl 17, 69117 Heidelberg, Germany \url{fengweilookuper@gmail.com}, \url{fengwei@mpia.de}}

\affil[2]{Universidad Nacional Aut\'onoma de M\'exico, Instituto de Radioastronom\'ia y Astrof\'isica, Antigua Carretera a P\'atzcuaro 8701, Ex-Hda. San Jos\'e de la Huerta,\\ 58089, Morelia, Michoac\'an, M\'exico}

\affil[3]{Center for Astrophysics $|$ Harvard \& Smithsonian, 60 Garden Street, Cambridge, MA 02138, USA}

\affil[4]{Laboratoire d'astrophysique de Bordeaux, Univ. Bordeaux, CNRS, B18N, allée Geoffroy Saint-Hilaire, 33615, Pessac, France}

\affil[5]{INAF-IAPS, Via Fosso del Cavaliere, 100, 00133 Rome, Italy}

\affil[6]{Dipartimento di Fisica, Università di Roma Tor Vergata, Via della Ricerca Scientifica 1, I-00133 Roma, Italy}

\abstract{
The origin of the stellar initial mass function (IMF) remains one of the central questions in astronomy. Nearly three decades ago, the resemblance between the core mass function (CMF) and the IMF inspired the community that stellar mass spectrum might be imprinted early in molecular-cloud cores and then mapped to the IMF through a simple efficiency factor. It has become gradually clear, however, that this apparent mapping involves multiple non-linear physical processes. Motivated by the spirit of the CMF2IMF conference at ESO Garching, this memo first reviews the historical quest to understand the origin of the IMF, and then sets the stage for building a shared understanding of current CMF measurements. We therefore compile several observational core catalogues at various environments and evolutionary stages into a common framework, implemented in the public Python package \texttt{CMF4All}. We show that the inferred high-mass CMF slope depends strongly on the adopted minimum fitting mass. Significantly steep slope is observed in the early-stage sample, indicating potentially evolving mass function in highest-mass slope. We conclude by outlining future directions to spare more efforts for both observers, numerical simulation and theoretical sides.
}

\keywords{interstellar medium - star formation - initial mass function}

\maketitle

\section{Introduction} \label{sec:intro}

The origin of the stellar initial mass function (IMF) remains one of the central questions in astronomy. Since the seminal work of \citet{salpeter1955}, the IMF has been recognized as the distribution of stellar birth masses and is commonly described by a power-law form at high masses,
\begin{equation} \label{eq:imf_alpha}
\frac{dN}{dM_\star} \propto M_\star^{-\alpha},
\end{equation}
or equivalently,
\begin{equation} \label{eq:imf_gamma}
\frac{dN}{d\log M_\star} \propto M_\star^{-\Gamma},
\end{equation}
where the two slope conventions are related by $\Gamma=\alpha-1$. In this notation, the Salpeter slope corresponds to $\alpha=2.35$, or $\Gamma=1.35$. At lower masses, the IMF flattens relative to the Salpeter power law and is commonly described by a broken power law or lognormal-like form \citep[e.g.,][]{miller&scalo1979, kroupa2001, chabrier2003}. Because stars of different masses dominate different channels of radiation, mechanical feedback, nucleosynthesis, and chemical enrichment, the IMF connects the small-scale physics of molecular-cloud fragmentation to the large-scale evolution of star clusters, galaxies, and the interstellar medium \citep[e.g.,][]{kennicutt1998, kroupa2002, bastian2010}. Its apparent regularity across many Galactic environments has long suggested that one or more robust physical processes may regulate the conversion of interstellar gas into stars, although possible environmental variations remain actively debated \citep[e.g.,][]{bastian2010, offner2014, krumholz2019, hennebelle2020, guszejnov2022}. Over the years, a major goal has therefore been to understand when and how the stellar mass distribution is first imprinted, as well as the physical processes into the play.

A particularly attractive possibility is that the IMF is inherited, at least partly, from the mass distribution of dense molecular cores, namely the core mass function (CMF). This idea was motivated by early millimeter and submillimeter continuum observations of star-forming regions, which found that the mass spectra of dense cores or condensations can display Salpeter-like high-mass slopes \citep[e.g.,][]{motte1998, testi1998, beuther2004}. With higher mass sensitivity, a CMF turnover mass is recovered to be shifted towards IMF with a constant conversion efficiency \citep[e.g., 30\% in][]{alves2007}. These results suggested that the stellar mass spectrum might already be partly encoded in the dense-gas structure before the final assembly of stars. 

\begin{figure}[!ht]
    \centering
    \includegraphics[width=0.7\linewidth]{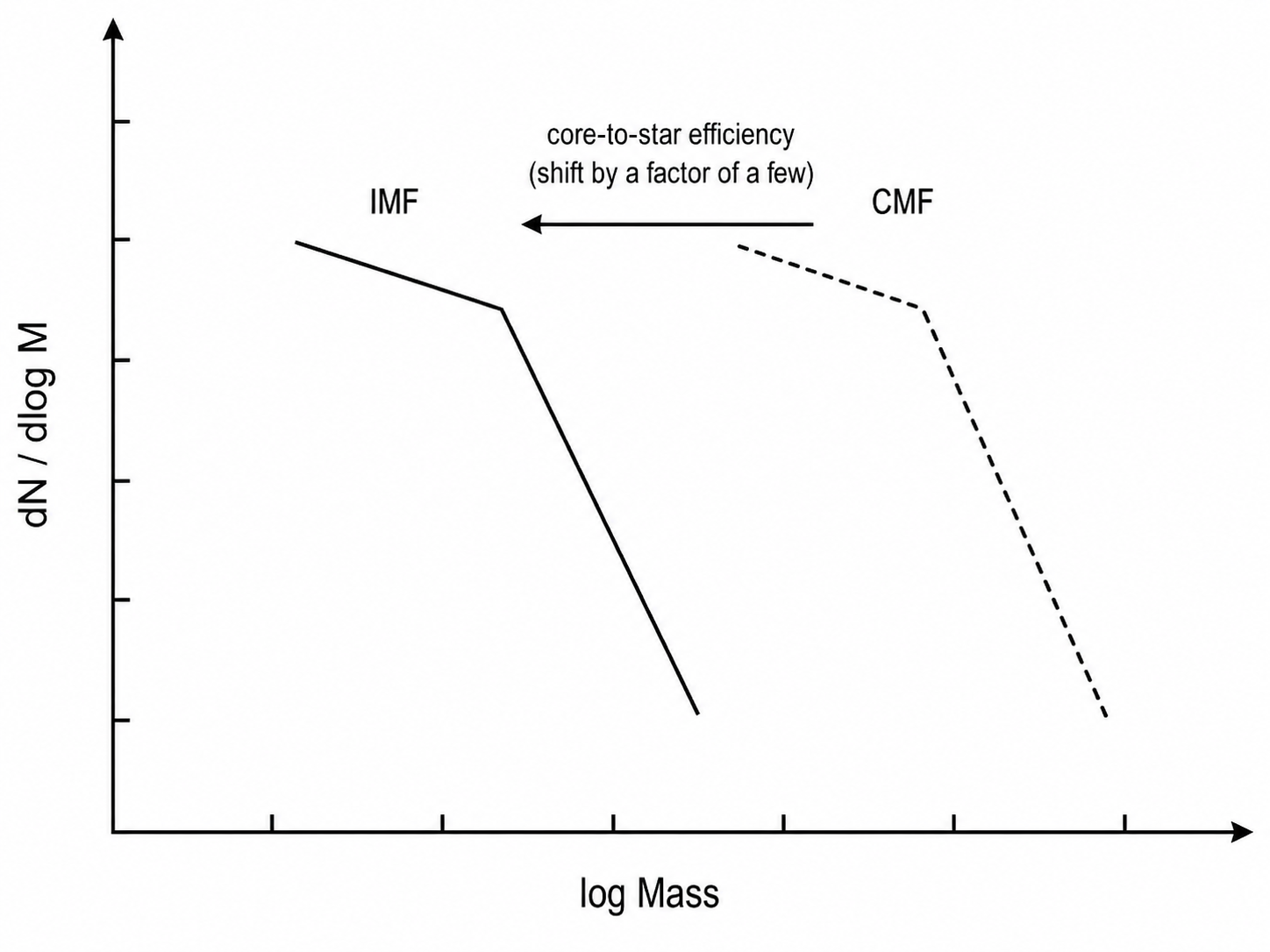}
    \caption{\textbf{Conceptual map from core mass function (CMF) to initial mass function (IMF)}.}
    \label{fig:cmf2imf}
\end{figure}

The launch of \textit{Herschel} enabled a great step forward by providing large, sensitive, and relatively uniform far-infrared surveys of nearby molecular clouds. In regions such as Aquila, hundreds of starless and prestellar cores were identified, allowing the CMF to be measured with much better statistics than before \citep[e.g.,][]{andre2010, konyves2015}. These surveys reinforced the empirical resemblance between the prestellar\footnote{In \textit{Herschel} studies, prestellar cores usually refer to starless dense cores, identified by the absence of compact $70~\mu$m emission, that are judged to be gravitationally bound, often through a Bonnor-Ebert stability criterion.} CMF and the IMF: the observed CMF was found to have a shape similar to the stellar IMF, but simply shifted toward higher masses by a factor of a few. This shift was often interpreted in terms of a simple core-to-star mapping:
\begin{equation}
M_\star \simeq \epsilon_{\rm core} M_{\rm core},
\end{equation}
where $\epsilon_{\rm core}$ is an effective core-to-star efficiency \citep[e.g.,][]{konyves2015}. The apparent resemblance between the CMF and IMF, therefore, offered an appealing interpretation: the stellar mass spectrum may be inherited from molecular-cloud fragmentation, with an effective mass conversion efficiency to shift the spectrum systematically. Such approximately uniform efficiency is thought to be through molecular outflows, for example. This mapping is illustrated in Fig.~\ref{fig:cmf2imf}.

However, nearby star-forming regions are not representative of the environments in which most stars, and especially massive stars, are thought to form. Establishing whether the same CMF–IMF connection holds across the broader diversity of Galactic star-forming environments, therefore, requires extending these studies to more distant, clustered, and massive regions. Massive stars rarely form in isolation \citep{lada2003}, and their birth environments are typically denser, more turbulent, more clustered, and more strongly affected by radiative and mechanical feedback than nearby low-mass clouds \citep{zinnecker2007, motte2018, beuther2025}. In such regions, the correspondence between a dense core and a final stellar system is expected to be less direct. A core may fragment into a multiple system, continue to accrete from a larger clump or filament, interact dynamically with neighbouring structures, or be affected by protostellar feedback while it is still being observed. Therefore, the CMF in massive protoclusters is not merely a high-mass extension of the nearby-cloud CMF, but rather a test of whether the same fragmentation and mass-assembly picture can operate under more extreme physical conditions.

High-resolution interferometric surveys have made this test observationally feasible. Especially with the Atacama Large Millimeter/submillimeter Array (ALMA), dense-core populations can now be resolved in distant Galactic massive clouds \citep[e.g.,][]{louvet2024, coletta2025, morii2026a} or even extragalactic clouds in the Large/Small Magellanic Cloud (LMC/SMC) \citep{traficante2026} at a few thousand astronomical units (au). Some of these studies report CMFs broadly compatible with a Salpeter-like high-mass slope, while others find significantly flatter, top-heavy, or otherwise environment-dependent distributions. When taking these results together, a fundamental question is raised: do the observed differences reflect genuine physical variations in the dense-core population, or do they arise from differences in resolution, spatial filtering, source extraction, temperature estimation, completeness limits, or even fitting methods?

This question is especially timely because the number of published CMF measurements is now large enough that the field has entered a comparative phase, but the comparison is never straightforward. Nearby \textit{Herschel} surveys and distant ALMA surveys often differ not only in sensitivity and resolution, but also in what they call a core, how they estimate core temperature, how they assign masses, how they define completeness, and how they measure the high-mass slope. A Salpeter-like or top-heavy CMF measured in one survey may therefore not be directly equivalent to the same quoted slope in another survey. Before the CMF can be understood as a possible bridge to the IMF, the existing measurements must first be placed into a common framework in which their assumptions and systematics are made explicit. 

This memo was motivated by the recent conference ``{\it CMF2IMF: The Origin of the Stellar Initial Mass Function}'', held at ESO Garching from 8–12 June 2026 \footnote{\url{https://www.eso.org/sci/meetings/2026/CMF2IMF.html}}. The meeting brought together observational and theoretical perspectives on the origin of the IMF and on the possible connection between core mass functions and stellar mass distributions. We had in-depth discussions during the conference and decide to write down the key inspirations as a reference for wider community. The main idea is not simply to ask whether the CMF is identical to the IMF, but to understand how dense cores, protostars, stellar systems, and the final IMF are connected through fragmentation, accretion, multiplicity, feedback, magnetic fields, turbulence, and environmental evolution. One lesson from the conference is that the CMF-IMF connection is no longer a simple, single-slope comparison problem. It is a problem of not just physics but also definitions, timescales, and even methodology. 

Motivated by this perspective, the goal of this memo is to compile the CMF measurements from various \textit{Herschel} and ALMA surveys into a common context for better comparisons and visualization. Here we assume all of the core mass measurements are reliable within their given uncertainties. We then compare the high-mass-end slopes across the different sample and with different fitting strategies. This comparison is intended as a practical starting point for aligning these CMF measurements and identifying which similarities among CMFs may be physically meaningful, which differences may be methodological, and what information is needed before CMF can be interpreted.

\section{Data} \label{sec:data}

The core catalogues used in this memo are compiled from a set of surveys discussed in the context of the CMF2IMF conference. These surveys cover a broad range of environments and evolutionary stages. The goal is not to define a complete census of all CMF studies (there are definitely many more than mentioned here), but to assemble a representative comparison set spanning nearby low-mass molecular clouds, Galactic-plane massive star-forming regions, the Galactic centre, and even extragalactic ALMA CMF measurements. This section briefly summarizes the main surveys included in the comparison and the role each plays in the broader CMF–IMF discussion. 

\begin{table*}[!ht]
\centering
\caption{
Summary of surveys and datasets considered in this memo.
$^{\mathrm{a}}$: The ALMA-IMF fields have various mass completeness from 0.80 to 1.64~\msun{}. For the purpose of CMF fitting, we take the conservative 1.64~\msun. 
$^{\mathrm{b}}$: The ALMAGAL survey calculates 0.13 and 0.37~\msun{} for near and far sample, and we take the conservative 0.37~\msun.
$^{\mathrm{c}}$: There are no completeness limit reported for the QUARKS sample, so we decide to estimate a rough value compared to the ALMAGAL survey based on their relative bandwidth and integration time. 
}
\label{tab:survey_summary}
\begin{tabular}{ccccc}
\hline
Survey name & Targets & Resolution & Cores & Completeness \\
 &  & [au] & number & [$M_{\odot}$] \\
\hline
HGBS / Aquila & Aquila complex & 4,700 & 685 & 0.30 \\
ALMA-IMF & Massive clouds & 2,700 & 597 & 1.64$^{\mathrm{a}}$ \\
ALMAGAL & Massive clumps & 2,000 & 6341 & 0.37$^{\mathrm{b}}$ \\
ASHES & Massive clumps & 5,000 & 715 & 1.20 \\
LANCET & Massive filament & 2,000 & 231 & 0.15 \\
QUARKS & Massive clumps & 2,000 & 3374  & 0.30$^{\mathrm{c}}$ \\
CMZ & Massive clouds & 2,000 & 808 & 2.0 \\
LMC / 30 Dor-10 & Massive clouds & 2,000 & 71 & 5.0 \\
\hline
\end{tabular}
\end{table*}

The nearby-cloud reference data set is provided by the Aquila cloud complex \citep{konyves2015} as part of the \textit{Herschel} Gould Belt Survey \citep{andre2010}. These observations identified large samples of starless and prestellar cores in nearby molecular clouds and established one of the clearest empirical cases for a prestellar CMF resembling the stellar IMF. Because the targets are nearby, the physical interpretation of individual prestellar cores is relatively mature compared with more distant massive regions. The Aquila CMF, therefore, serves as an important benchmark for the classical CMF–IMF picture.

The ALMA-IMF Large Program \citep{motte2022} extends the CMF question to massive protoclusters in the Milky Way. ALMA-IMF observed 15 massive clouds spanning a wide range of evolutionary stages at a matched physical resolution of $\sim2{,}700$~au, mosaicking areas of about 1 to 3~pc in size. The explicit goal is to understand the origin of the stellar IMF through their precursor core populations. The survey provides one of the largest and most homogeneous ALMA-based core samples in massive protoclusters. In its CMF analysis, ALMA-IMF extracted several hundred gravitationally bound cores and reported a high-mass CMF slope that is flatter than the Salpeter IMF slope \citep[e.g.,][]{louvet2024}. They also explored the environmental effects or evolution in a single cloud \citep{pouteau2022} as well as prestellar and protostellar core differences \citep{nony2023}. These make ALMA-IMF once a key reference for assessing whether massive clustered environments produce CMFs that are different from the nearby-cloud case.

ALMAGAL Large Program \citep{molinari2025} provides a complementary Galactic-wide perspective. Rather than focusing on a small number of selected protoclusters, ALMAGAL targets the largest unbiased sample of 1013 massive clumps across the Milky Way. The ALMAGAL data set enables statistical studies of clump-to-core fragmentation over a wide range of Galactic environments and evolutionary stages. Its unprecedented large sample size makes it especially valuable for exploring how core demographics, fragmentation properties, and CMF measurements vary across the broader high-mass star-formation population \citep[e.g.,][]{coletta2025}.

ASHES, the ALMA Survey of $70~\mu{m}$ Dark High-mass Clumps in Early Stages, focuses on cold and $70~\mu{m}$-dark massive clumps that are expected to represent the earliest stages of high-mass star formation \citep{sanhueza2019, morii2023}. By targeting 39 $70~\mu{m}$-dark infrared dark clouds, ASHES provides a view of dense-core populations before strong protostellar heating and feedback dominate the environment. Its CMF measurements are therefore particularly important for asking whether the high-mass CMF is already established at the prestellar or very early protostellar stage, or whether it evolves as cores grow through accretion and become increasingly affected by feedback \citep{morii2026a}. Another feature of the ASHES survey is that cores are characterized by their protostellar activity and gravitational boundness, which enables to identify robust prestellar cores \citep{li2023, morii2024}.

The LANCET survey covers the whole 14-parsec, nearly linear G316.8 filament, which is located on the near side of the Scutum-Centaurus Arm \citep{xu2026a}. Compared to the Galactic-wide sample, the G316.8 filament itself provides an excellent controlled case: it comprises three contiguous subregions with comparable molecular gas reservoirs ($10{,}000~M_{\odot}$) yet spanning a clear evolutionary sequence from a northern infrared dark cloud (young) through a central massive young stellar object (intermediate) to a southern \ion{H}{ii} region (evolved). Its unique value to CMF studies is its {\it in-situ} evolutionary sequence by controlling distance and global environment (metallicity and Galactic dynamics). By comparing CMFs among the young, intermediate, and evolved subregions, LANCET provides a framework for testing whether the CMF changes with local evolutionary stage and feedback activity within the same parent cloud.

The QUARKS survey observed 139 massive star-forming clumps \citep{liu2024, xu2024}. It uses one or two pointings toward the millimeter peaks of massive clumps. QUARKS is more biased toward active, evolved star-forming regions with $L/M \geq 4~L_\odot/M_\odot$ \citep{xu2024}. In this sense, QUARKS provides a natural complement to early-stage surveys such as ASHES, which focus on cold and infrared-dark massive clumps. Although the QUARKS sample is about an order of magnitude smaller than ALMAGAL, its observations are somewhat deeper for individual fields. This is partly due to its broader spectral coverage and, on average, longer integration time per pointing. The depth of QUARKS is also reflected in the number of detected compact cores; despite the much smaller number of target fields, the total number of cores identified in QUARKS (Jiao et al. in prep.) is half of that reported by ALMAGAL. 

We also include a recent ALMA mini-survey of three massive clouds in the Galactic Central Molecular Zone (CMZ): the 20~km~s$^{-1}$ cloud, Sgr~C, and dust ridge cloud~e \citep{lu2020, zhang2025, xu2025}. The CMZ is a unique star-forming environment compared to typical Galactic disk clouds. It is also a local analogue of more distant galactic nuclei. These conditions include strong magnetic fields, intense turbulence, elevated cosmic-ray ionization rates, large external pressure, strong tidal forces and gravitational shear, and feedback associated with the Galactic centre environment, including the supermassive black hole Sgr~A* \citep[e.g.,][]{henshaw2023}. Because these clouds are all located in the inner 300~pc molecular zone of the Milky Way, their distances are effectively similar, allowing their core populations to be compared at comparable physical resolutions of the order of $2{,}000$~au. Another important advantage of these surveys is their large mosaic coverage, designed to follow the footprint of the extended dust emission rather than only the brightest compact peaks. The CMZ cores provide an important high-pressure, high-turbulence comparison set for testing whether CMFs measured in extreme Galactic environments differ from those in nearby clouds and more typical massive star-forming regions in the Galactic disk.

Finally, the recent ALMA study of 30 Dor-10 in the LMC by \citet{traficante2026} provides the first measurement of the CMF in an external galaxy at a physical resolution of about $2{,}000$~au. This work extends the CMF discussion beyond the Galactic plane. Putting this work into our framework provides a crucial test of whether the fragmentation properties and core mass distribution inferred in the Milky Way also apply in a low-metallicity environment, or, broadly speaking, the early universe. 

\section{Method} \label{sec:method}

\subsection{Mass-function representation}

All CMFs in this memo are expressed using the same slope conventions as introduced in Eqs.~(\ref{eq:imf_alpha}) and (\ref{eq:imf_gamma}). In brief, we use $\alpha$ for the linear differential form, $dN/dM \propto M^{-\alpha}$, and $\Gamma$ for the logarithmic form, $dN/d\log M \propto M^{-\Gamma}$, with $\Gamma=\alpha-1$. For each survey, we construct both the differential CMF and the complementary cumulative CMF. 

The differential CMF is computed in logarithmic mass bins and is written as
\begin{equation} \label{eq:discrete}
\left(\frac{dN}{d\log M}\right)_i
= \frac{N_i}{\Delta \log M},
\end{equation}
where $N_i$ is the number of cores in the $i$-th logarithmic mass bin, and $\Delta \log M$ is the adopted bin width. The uncertainty in each bin is estimated from Poisson counting statistics,
\begin{equation}
\sigma_i = \frac{\sqrt{N_i}}{\Delta \log M}.
\end{equation}
The binned differential CMF provides an intuitive visualization of the mass distribution, but it is sensitive to the adopted bin edges and bin width.

We also use the complementary cumulative CMF,
\begin{equation}
N(\geq M),
\end{equation}
or, equivalently, its normalized form $N(\geq M)/N_{\rm tot}$. The cumulative form avoids binning and is useful for visualizing the high-mass tail. However, neighbouring cumulative points are not statistically independent. For this reason, we use the differential and cumulative forms just for visualization, while the power-law slope is derived from the method without binning mass distribution as explained below.

\subsection{Power-law fitting}

The high-mass tail of each CMF is fitted with a single power law above a minimum fitting mass ($M_{\min}$). For masses $M_i \geq M_{\min}$, the probability density is written as
\begin{equation}
p(M|\alpha) =
\frac{\alpha-1}{M_{\min}}
\left(\frac{M}{M_{\min}}\right)^{-\alpha},
\qquad M \geq M_{\min}.
\end{equation}
The corresponding maximum-likelihood estimator \citep[MLE;][]{clauset2009} is
\begin{equation}
\alpha = 1 + N
\left[
\sum_{i=1}^{N}
\ln\left(\frac{M_i}{M_{\min}}\right)
\right]^{-1},
\end{equation}
where $N$ is the number of cores with $M_i \geq M_{\min}$. The logarithmic slope is then obtained from $\Gamma=\alpha-1$. The approximate analytic uncertainty is
\begin{equation}
\sigma_\alpha \simeq \frac{\alpha-1}{\sqrt{N}},
\end{equation}
which is also the uncertainty in $\Gamma$.

Throughout this memo, the unbinned maximum-likelihood approach is adopted because it avoids the dependence of the fitted slope on arbitrary binning choices. The binned differential CMF will still be shown along, but the quoted power-law slopes are derived from the unbinned high-mass tail. 

\subsection{Choice of minimum fitting mass}

The choice of $M_{\min}$ can substantially affect the inferred CMF slope and is arguably one of the most delicate methodological choices in published CMF studies. In principle, two choices are commonly used, each with its own implicit assumptions. The first is to adopt the observational completeness limit, $M_{\rm comp}$, as the minimum fitting mass. This completeness limit is often defined as the core mass above which the recovery rate\footnote{Typically, artificial cores are injected into the noisy image, and the source-extraction procedure is repeated to determine what fraction of injected cores can be recovered.} exceeds 90\%. Fitting a power law from $M_{\rm comp}$ implicitly assumes that the mass distribution already follows a power-law form above this limit. If this assumption is not valid, the fitted slope becomes a mathematical description of the overall decreasing trend rather than a physically meaningful measurement of the high-mass tail.

This issue is particularly important if the mass function is not described by a single power law. For example, based on Gaia DR3, \citet{wang2026} suggested that the stellar IMF within $0.25{-}1~\msun$ contain two power-law segments with a break point at $0.40^{+0.01}_{-0.01}$~\msun. At smaller masses, a shallower slope of $\alpha_1=0.75$ is observed, and toward the high-mass end, a steeper slope $\alpha_2=2.07$ (Salpeter-like) is recovered. If some aspects of the IMF are inherited from the CMF, then similar broken or multi-segment power-law behaviour may also appear in the CMF above $M_{\rm comp}$. Indeed, several CMFs in Fig.~\ref{fig:cmf_mcomp_fit} show a shallow component immediately above the completeness limit, followed by a steeper decline at higher masses. The two local minima in the KS-distance curve shown in Fig.~\ref{fig:ks-show} may also indicate the presence of two approximate power-law regimes. In such cases, fitting the full mass range above $M_{\rm comp}$ with a single power law yields an average slope over multiple regimes, rather than the slope of the true high-mass tail.

The second choice is to determine $M_{\min}$ statistically from the observed mass distribution. In this approach, a set of candidate $M_{\min}$ thresholds is tested. For each candidate threshold, a power law is fitted to the subset of cores above that mass, and the Kolmogorov--Smirnov (KS) distance is computed between the empirical cumulative distribution of the fitted tail and the best-fitting power-law model. The KS distance is defined as
\begin{equation} \label{eq:ks}
D_{\rm KS}=\sup_{M\ge M_{\min}} \left|F_{\rm emp}(M)-F_{\rm model}(M)\right|,
\end{equation}
where $F_{\rm emp}(M)$ is the empirical cumulative distribution of the data and $F_{\rm model}(M)$ is the cumulative distribution predicted by the fitted power law. The preferred $M_{\min}$ is then chosen as the value that minimizes $D_{\rm KS}$. Compared with a completeness-based threshold, the KS-selected threshold is designed to identify the mass range over which the observed distribution is most robustly described by a single power law. It therefore asks a different question: above what mass is the observed CMF best represented by a power-law tail?


The two fitting thresholds can differ substantially, and this difference can lead to different fitted slopes even for the same core catalogue. We therefore emphasize that a slope measured above $M_{\rm comp}$ and a slope measured above a KS-selected $M_{\min}$ are not necessarily equivalent. If the goal is to describe the mass distribution over the largest observationally reliable sample, then adopting $M_{\rm comp}$ is appropriate, because it includes all cores above the completeness limit. Another merit of completeness case is to use as many as reliable data points. It may reduce the statistical uncertainty. However, if the goal is to measure the absolute slope of the high-mass power-law tail, for example to compare with a Salpeter-like slope, then a statistically selected threshold such as $M_{\rm KS}$ is more appropriate. In this sense, the completeness-based choice is tied to observational reliability, whereas the KS-based choice is tied to the validity of the power-law description. 

In this memo, we test two cases of $M_{\rm min}$ for the retrieved core catalogues in Table~\ref{tab:survey_summary}. This procedure is implemented in the Python package \texttt{CMF4All} \footnote{\url{https://github.com/XFengwei/CMF4All}}.

\section{Results} \label{sec:result}

\subsection{Case A: completeness mass as minimum fitting mass}

\begin{figure}[!ht]
\centering
\includegraphics[height=0.37\linewidth]{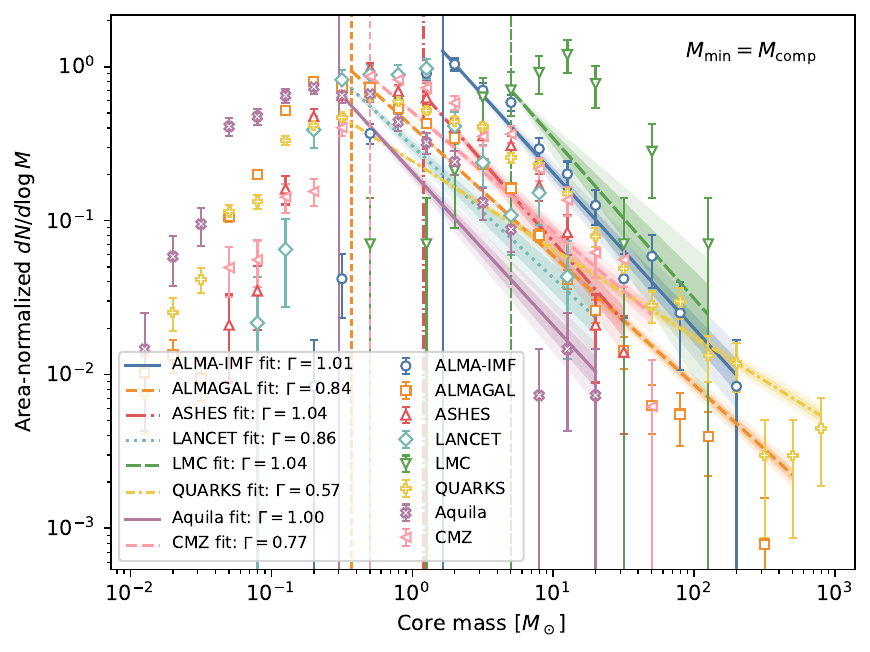}
\includegraphics[height=0.37\linewidth]{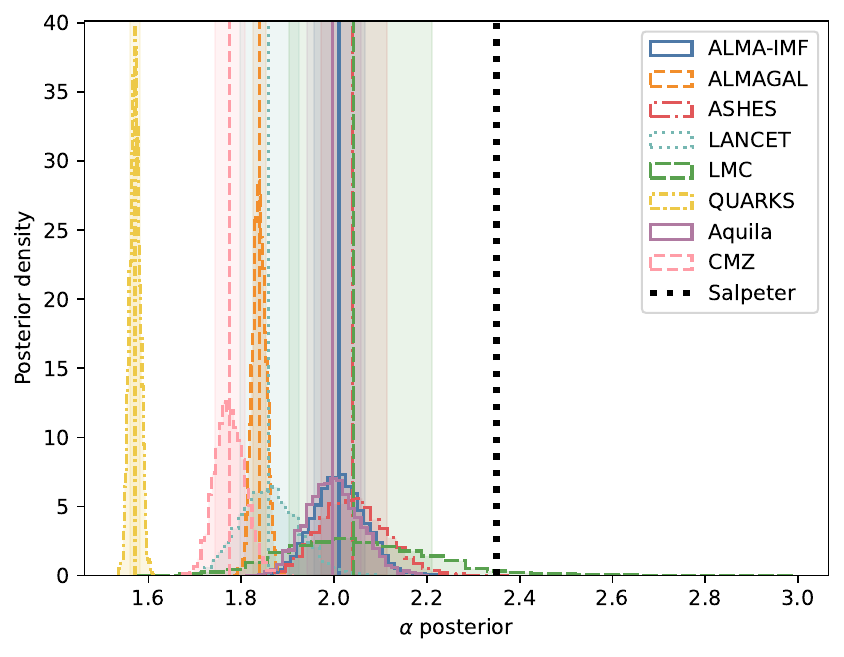}
\caption{
\textbf{Completeness-based CMF fits and posterior distributions of the fitted slopes.}
Top: Differential CMFs for the surveys included in this comparison. For each survey, the high-mass tail is fitted with a single power law above the adopted completeness mass, $M_{\min}=M_{\rm comp}$. The slopes shown for the CMF fits are expressed in the logarithmic convention, $dN/d\log M \propto M^{-\Gamma}$.
Bottom: Posterior distributions of the corresponding linear power-law slope, $\alpha=\Gamma+1$, shown as colour-matched histograms. Each posterior is generated from 10,000 Monte-Carlo realizations, and the shaded region marks the median value with the 16th--84th percentile interval. The black dashed vertical line indicates the Salpeter slope, $\alpha=2.35$. This figure illustrates the survey-to-survey variation obtained when the minimum fitting mass is tied to each survey's completeness limit.
}
\label{fig:cmf_mcomp_fit}
\end{figure}

As shown in Fig.~\ref{fig:cmf_mcomp_fit} and listed in Table~\ref{tab:fit_summary}, when the minimum fitting mass is fixed to the reported completeness mass, $M_{\min}=M_{\rm comp}$, all of the surveys yield power-law slopes that are flatter than the Salpeter value. At face value, this would suggest that the majority of the compiled CMFs are top-heavy relative to the stellar IMF. However, the differential CMFs in Fig.~\ref{fig:cmf_mcomp_fit} clearly demonstrate that the completeness-based fits are often not ideal descriptions of the observed mass distributions. In many cases, the CMF immediately above $M_{\rm comp}$ does not behave as a single declining power law. Instead, it shows a shallow decrease, a nearly flat segment, or even a mild rise before transitioning to a steeper decline at higher masses. Fitting one power law over this entire range therefore averages over both the shallow intermediate-mass part and the steeper high-mass tail. 

Note that here we didn't consider any classification of cores (i.e., prestellar or protostellar). This can make difference between our obtained slope and that reported in the literature. \citet{konyves2015} found Salpeter slopes for prestellar and for starless cores in Aquila, while here we have top-heavy for the total core sample. 

This behaviour has a substantial consequence for the inferred slope. If the true high-mass tail begins only above a mass scale larger than $M_{\rm comp}$, then adopting $M_{\rm comp}$ as the fitting threshold includes cores that belong to the turnover or transition region of the CMF. The resulting power-law fit is then pulled toward a flatter slope, producing an apparently top-heavy CMF even when the highest-mass tail may be much steeper. This effect is reflected in the bottom panel of Fig.~\ref{fig:cmf_mcomp_fit}, where the posterior distributions of $\alpha$ are significantly below the Salpeter value. 

\begin{table*}[!ht]
\centering
\caption{
{\bf Summary of fitting results}.
The subscript of ``comp'' represents the case when the minimum fitting mass is determined by completeness limit; ``KS'' is the case determined by the KS-distance criteria.
}
\label{tab:fit_summary}
\begin{tabular}{cccccc}
\hline
Survey name & $N_{\rm comp}$ & $\alpha_{\rm comp}$ & $M_{\rm KS}$ & $N_{\rm KS}$ & $\alpha_{\rm KS}$ \\
 & & & [\msun] & \\
\hline
HGBS / Aquila & 318 & $2.00(0.05)$ & 0.4 & 262 & $2.10(0.07)$ \\
ALMA-IMF & 352 & $2.01(0.05)$ & 3.8 & 167 & $2.22(0.10)$ \\
ALMAGAL  & 3485 & $1.84(0.01)$ & 3.9 & 451 & $2.35(0.07)$ \\
ASHES & 242 & $2.04(0.07)$ & 7.2 & 34  & $\mathbf{3.30(0.39)}$ \\
LANCET & 201 & $1.86(0.06)$ & 1.0 & 94 & $2.45(0.16)$ \\
QUARKS & 2539 & $1.57(0.01)$ & 5.9 & 443 & $2.03(0.05)$ \\
CMZ & 613 & $1.77(0.03)$ & 5.3 & 104 & $2.58(0.17)$ \\
LMC / 30 Dor-10 & 54 & $2.04(0.15)$ & 9.6 & 37 & $2.66(0.29)$ \\
\hline
\end{tabular}
\end{table*}

The systematic appearance of shallow completeness-based slopes therefore should not be interpreted as direct evidence that all surveyed regions have intrinsically top-heavy core populations. Rather, Fig.~\ref{fig:cmf_mcomp_fit} demonstrates that the CMF shape above the completeness limit is often more complex than a single power law. The mass range between $M_{\rm comp}$ and the onset of the steeper high-mass tail may contain information about a second shallower power law, the CMF turnover, or the transition between incomplete and complete sampling of the dense-core population.

\subsection{Case B: KS-determined minimum fitting mass}

Unlike the completeness-based case, KS-determined minimum fitting mass, $M_{\min}=M_{\rm KS}$ threshold is self-adaptively selected from the observed mass distribution as the point above which a single power law provides the best description of the high-mass tail. For each candidate $M_{\min}$, we fit unbinned core masses above that threshold and compute the KS-distance $D_{\rm KS}$ as defined by Eq.~(\ref{eq:ks}) between the empirical cumulative distribution of the fitted sample and the best-fitting power-law model. 

\begin{figure}[!ht]
\centering
\includegraphics[width=0.5\linewidth]{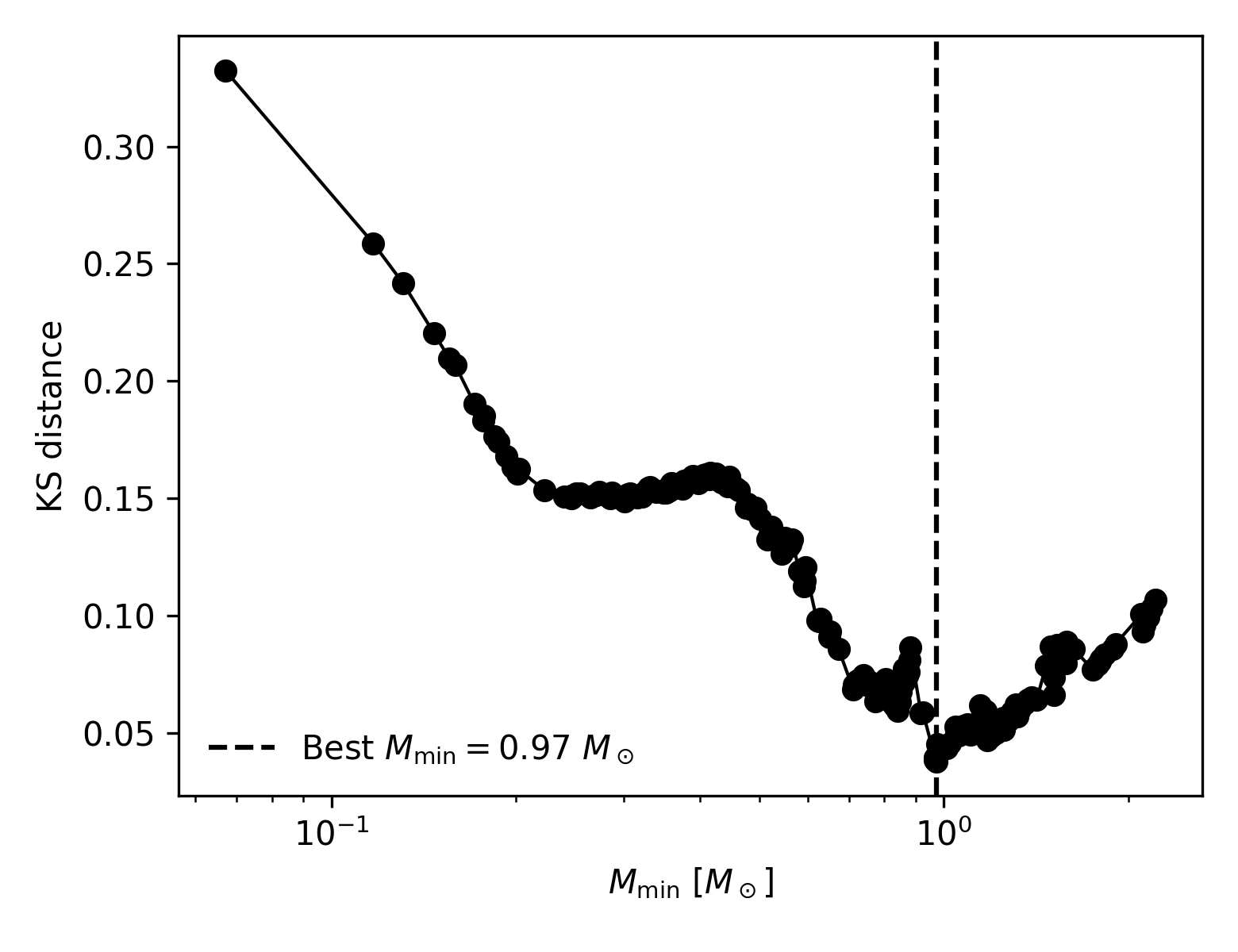}
\caption{
\textbf{Selection of the minimum fitting mass using the KS distance.} Here shows an example of the LANCET survey core sample. The KS distance is shown as a function of the candidate minimum fitting mass, $M_{\min}$, for the LANCET core catalogue. For each trial $M_{\min}$, a single power law is fitted to the unbinned core masses above that threshold, and $D_{\rm KS}$ is computed between the empirical cumulative distribution of the fitted tail and the best-fitting power-law model. The minimum of the curve defines the KS-selected minimum fitting mass, $M_{\rm KS}$, used in the uniform fitting procedure.
}
\label{fig:ks-show}
\end{figure}

Fig.~\ref{fig:ks-show} gives an exemplar $D_{\rm KS}$ curve for the LANCET core catalogue as an illustrative example. Above the completeness limit of $0.15~\msun$, the curve shows two local minima. Comparing this behaviour with the differential CMF in Fig.~\ref{fig:cmf_mcomp_fit}, these two minima correspond to two starting masses for approximate power-law regimes. The lower minimum corresponds to the onset of a shallower component, while the higher minimum marks the beginning of the steeper high-mass tail. This demonstrates that, if a single power law is fitted from a mass lower than $0.97~\msun$, the fitted sample may include more than one power-law component, and the resulting slope will not accurately represent the true high-mass tail. In this sense, the KS-distance curve provides a useful diagnostic for identifying where the CMF becomes well described by a single power law.

\begin{figure}[!ht]
\centering
\includegraphics[height=0.37\linewidth]{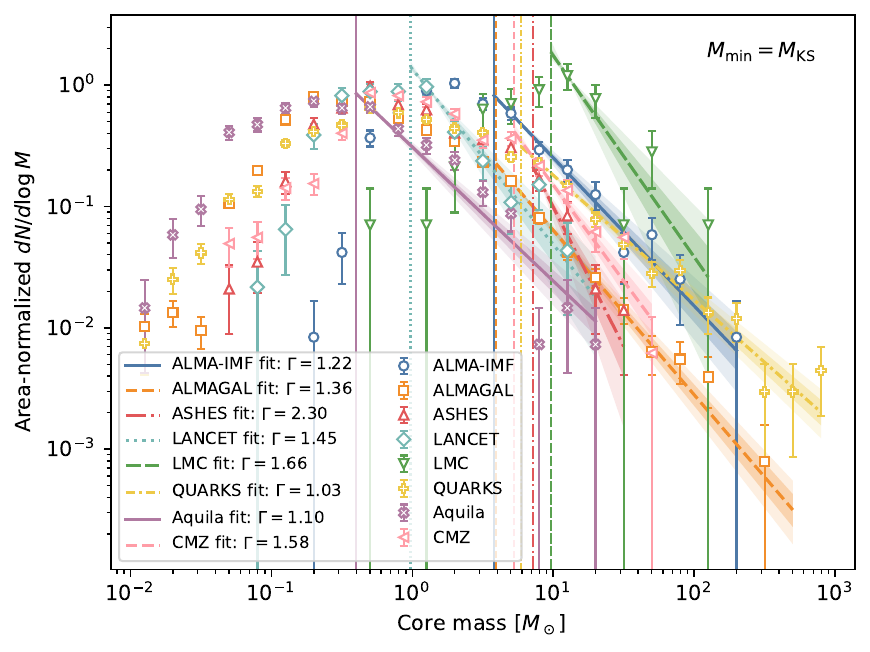}
\includegraphics[height=0.37\linewidth]{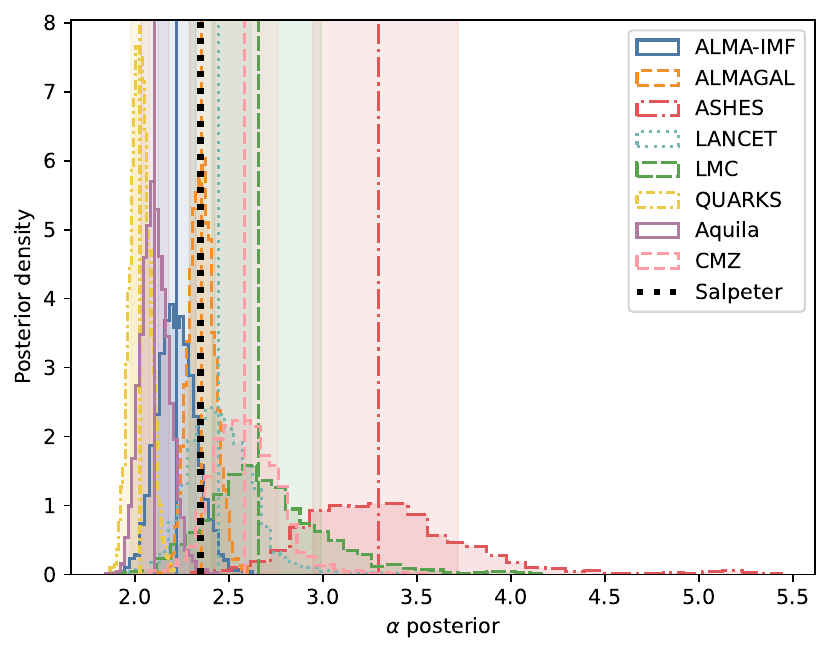}
\caption{
\textbf{KS-based CMF fits and posterior distributions of the fitted slopes.}
Top: Differential CMFs for the surveys included in this comparison. For each survey, the high-mass tail is fitted with a single power law above the KS-selected minimum fitting mass, $M_{\min}=M_{\rm KS}$. The slopes shown for the CMF fits are expressed in the logarithmic convention, $dN/d\log M \propto M^{-\Gamma}$.
Bottom: Posterior distributions of the corresponding linear power-law slope, $\alpha=\Gamma+1$, shown as colour-matched histograms. Each posterior is generated from 10,000 Monte-Carlo realizations, and the shaded region marks the median value with the 16th--84th percentile interval. The black dashed vertical line indicates the Salpeter slope, $\alpha=2.35$. This figure illustrates the survey-to-survey variation obtained when the minimum fitting mass is tied to each survey's $M_{\rm KS}$.
}
\label{fig:cmf_mks_fit}
\end{figure}

For all surveys, the obtained $M_{\rm KS}$, the corresponding number of fitted cores, $N_{\rm KS}$, and the fitted slope, $\alpha_{\rm KS}$, are listed in Table~\ref{tab:fit_summary}. The corresponding fits are shown in Fig.~\ref{fig:cmf_mks_fit}. The resulting slopes are systematically different from those obtained in Case~A. Once the fit is restricted to the KS-selected power-law tail, all surveys have steeper slopes and move closer to the Salpeter value. This indicates that the apparently top-heavy slopes in Case~A are not necessarily a robust property of the highest-mass end. Instead, they are largely driven by including the monotonic but non-single-power-law part of the CMF immediately above the completeness limit, where the distribution is not yet well described by a single power law. In other words, many CMFs appear top-heavy when fitted from $M_{\rm comp}$, but become approximately Salpeter-like when the fitting range is restricted to the mass interval that is statistically consistent with a single power law.

Although the KS-selected fits generally steepen the slopes, not all of them are statistically consistent with a Salpeter-like slope once their uncertainties are considered. In some surveys, the fitted high-mass tail remains significantly flatter or steeper than Salpeter, while in others the uncertainty is large because only a limited number of cores remain above $M_{\rm KS}$. For example, the slope in ALMA-IMF remains top-heavy, which is consistent with result in \citet{louvet2024}. The top-heavy CMF is also prominent in the QUARKS and Aquila surveys. We note here that the KS-based analysis does not imply that CMFs share more Salpeter-like high-mass slope. Rather, it shows that part of the apparent top-heavy behaviour in the completeness-based fits is methodological, while the remaining survey-to-survey differences may still reflect sample size, source definition, environment, or genuine physical variation.

An especially interesting exception is the ASHES survey. Its CMF shows evidence for two power-law-like regimes above the completeness limit. The first, shallower segment extends from $\sim0.4~\msun$ to the KS-selected lower fitting mass, $M_{\rm KS}=7.2~\msun$, while the second segment continues above $M_{\rm KS}$. This behaviour suggests that the mass range immediately above the completeness limit is not part of the same high-mass tail as the most massive cores. Instead, ASHES appears to contain a shallower intermediate-mass component followed by a much steeper high-mass decline. The slope of this second power-law tail is noticeably steeper than the Salpeter value. As shown by the green points and fitted line in Fig.~\ref{fig:cmf_mks_fit}, ASHES has a fitted slope of $\Gamma=2.30$, corresponding to $\alpha=3.30$, although with a relatively large uncertainty due to the limited number of cores in the fitted tail. This top-light behaviour at the highest-mass end may be physically interesting because ASHES targets very early-stage, $70~\mu$m-dark massive clumps with globally low luminosity-to-mass ratios, $L/M<1$. It may therefore probe dense-core populations before substantial protostellar growth has populated the most massive end of the CMF.

\section{Discussion}\label{sec:discuss}

\subsection{Method-dependent CMF slope}

One central result of this comparison is that the inferred high-mass CMF slope depends sensitively on how the minimum fitting mass is chosen. When the power-law fit is performed above the survey completeness mass, $M_{\min}=M_{\rm comp}$, many CMFs appear flatter than Salpeter and would therefore be classified as top-heavy. However, the differential CMFs show that the mass range immediately above $M_{\rm comp}$ is often not a clean high-mass power-law tail. Instead, it frequently contains a shallow or flattened component, as seen in ALMAGAL, ASHES, LANCET, and the CMZ sample, or even a mildly rising component, as seen in the LMC case, before the distribution steepens at higher masses. A single power law fitted across this full range therefore averages over both the lower-mass component and the steeper high-mass tail, producing an artificially shallow slope.

A clear example of this effect is shown in Fig.~5 of \citet{louvet2024}, where the ALMA-IMF team demonstrated how the fitted CMF slope varies as a function of $M_{\min}$. Their analysis shows that the inferred slope can become steeper as $M_{\min}$ increases, although the top-heavy nature of the ALMA-IMF CMF remains robust across the full sample. This test illustrates an important point: even when the qualitative conclusion of a study is stable, the absolute value of the fitted slope can depend strongly on the adopted minimum fitting mass. Therefore, any CMF study for which the absolute slope is central to the scientific interpretation should explicitly test how the result changes with $M_{\min}$.

This does not mean that completeness-based fitting is wrong. The completeness mass answers an observationally straightforward question: what is the behaviour of the core mass distribution above the minimum mass at which the catalogue is considered reliable? However, $M_{\rm comp}$ does not necessarily mark the onset of a single power-law regime. Conversely, a statistically selected threshold such as $M_{\rm KS}$ is designed to identify the mass range over which a power law provides an adequate description, but it may discard complete and physically meaningful cores below that threshold. These two choices therefore answer different questions. For cross-survey comparisons, the most important requirement is not that every study adopt the same preferred threshold, but that the fitting threshold and its justification be reported clearly and, where possible, tested under a consistent prescription.

A second take-home message is that, like the IMF, the CMF may not be adequately described by a single power law over the full mass range above the completeness limit. Instead, the CMF may be better represented by a broken power law, a lognormal-like distribution extending to high masses, a lognormal component with an additional power-law tail, or a distribution with an explicit high-mass cutoff. Distinguishing among these functional forms is itself non-trivial. For example, the ALMA-IMF analysis explored statistical comparisons between lognormal and power-law descriptions, but the available core sample was not yet large enough to decisively distinguish between them \citep[see discussion in][]{louvet2024}. This issue becomes particularly important for the new generation of interferometric surveys, which can reach mass sensitivities of order $0.1~\msun$ and therefore probe well below the lower-mass end of the putative high-mass power-law tail. In this regime, fitting a single power law above $M_{\rm comp}$ can mix the turnover, curved or shallow components, and the high-mass tail into one slope. This is closely related to one of the open questions discussed at the CMF2IMF conference: what functional form should be used to describe the CMF? Future CMF studies should therefore report both the observational completeness limit and the minimum mass adopted for power-law fitting, and should state explicitly whether the fitted slope describes all complete cores or only the statistically selected high-mass tail.

\begin{figure}[!ht]
    \centering
    \includegraphics[width=0.8\linewidth]{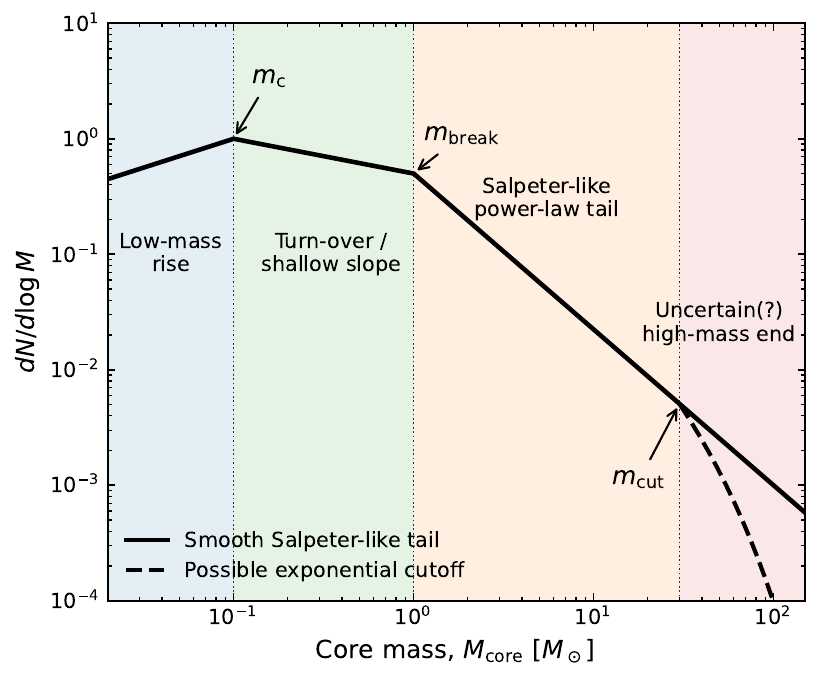}
    \caption{\textbf{Conceptual core mass function (CMF).}
    Schematic illustration of a segmented CMF in the logarithmic mass-function form. The lowest-mass regime rises from the proto-brown-dwarf scale toward a characteristic mass, $m_{\rm c}\sim0.1~\msun$, where the distribution turns over. Above $m_{\rm c}$, the CMF follows a shallow declining segment until a break mass of order $m_{\rm break}\sim1~\msun$. From $\sim1~\msun$ to several tens of $\msun$, the CMF may enter a canonical high-mass regime with a Salpeter-like slope. The behaviour at the highest-mass end remains uncertain: the distribution may continue as a smooth power-law tail, or it may show an exponential-like cutoff if the most massive cores are limited by cloud mass, evolutionary stage, feedback, or small-number sampling. This schematic emphasizes that a single fitted slope may mix several physically or observationally distinct mass regimes.
}

    \label{fig:cmf_segmentation}
\end{figure}

A useful way forward is to treat the CMF as a segmented distribution rather than forcing it into a single descriptive slope. As illustrated schematically in Fig.~\ref{fig:cmf_segmentation}, different mass ranges may encode different physical or observational regimes: the lowest-mass end is shaped by sensitivity, completeness, and the onset of core identification (see \citealt{palau2024} for a review on these limitations); the characteristic mass or turnover may reflect the dominant fragmentation scale; the intermediate-mass part may form a shallow component; and only the upper tail may approach a Salpeter-like or steeper power-law behaviour. At the highest masses, an additional uncertainty remains: the distribution may continue as a smooth power law, or it may show an exponential-like cutoff if the most massive cores are limited by cloud mass, evolutionary stage, feedback, or small-number statistics. Observationally testing this full structure requires progress at both ends of the CMF. On the low-mass side, higher sensitivity and more robust completeness tests are needed to determine whether the turnover and shallow component are real or shaped by source extraction and noise. On the high-mass side, improved mass estimates for the most massive and compact cores are essential, including better constraints on temperature, optical depth, dust opacity, free-free contamination, and multiplicity. Only by extending the dynamic range in both directions can future surveys determine whether the CMF is best described by a lognormal-like form, a broken power law, a Salpeter-like tail, or a high-mass cutoff.

\subsection{Toward an evolutionary view of the CMF}

The KS-based comparison suggests that the CMF may contain evolutionary information, although this interpretation must be made with caution. ASHES remains among the steepest CMFs in our comparison and appears top-light relative to most other surveys. Since ASHES targets 70~$\mu$m-dark massive clumps with very low luminosity-to-mass ratios, $L/M<1$, it likely probes an earlier stage of massive star formation than many of the other ALMA samples considered here. A natural interpretation is that the high-mass end of the CMF has not yet been fully populated in such early regions. However, the KS-selected minimum fitting masses, $M_{\rm KS}$, differ from survey to survey, and therefore the fitted slopes may not always probe exactly the same mass regime. The top-light behaviour of ASHES should thus be viewed as suggestive rather than conclusive: it may indicate an early high-mass cutoff, but it may also partly reflect the mass interval selected as the statistically preferred power-law tail.

With this caveat in mind, the ASHES result fits naturally into a picture in which the CMF is a present-day mass function rather than a fixed initial condition. In young regions, many low- and intermediate-mass condensations may already be present, while the most massive cores remain rare because they require additional time to grow through accretion, mass concentration, or hierarchical assembly within the protocluster potential. In this view, a top-light CMF is not simply a different slope, but a mass function observed before the high-mass end has been fully assembled. A similar idea appears in simulations of forming stellar populations: in the STARFORGE simulations, \citet{grudic2023} found a high-mass cutoff rather than a fully sampled Salpeter-like tail, with massive stars tending to begin accreting earlier and continue accreting longer than average stars. The analogy for the CMF is that a mass function observed during assembly need not yet sample its final high-mass tail.

Observational comparisons between prestellar and protostellar CMFs also support this evolutionary view. Studies such as \citet{nony2023} and \citet{morii2026a} suggest that prestellar CMFs in high-mass star-forming regions can remain broadly Salpeter-like, whereas the most massive cores contributing to a top-heavy high-mass tail are predominantly protostellar. This distinction implies that the massive end of the CMF is unlikely to be set by fragmentation alone. It must also reflect time-dependent mass growth, continued accretion from the surrounding clump or filaments, and the changing visibility of cores as they become internally heated. Statistical studies of massive protoclusters point in the same direction: protocluster members appear to grow in mass and density, become more closely spaced, and show increasing central concentration or primordial mass segregation as evolution proceeds \citep{xu2024, coletta2025, elia2026, schisano2026}. If the CMF evolves in this way, then the high-mass end should be progressively built up rather than fixed at the earliest observable stage \citep[see also][]{pouteau2022}. 

With more and better CMF measurements in single star-forming regions, it becomes possible that people can compare CMF measurements across different evolutionary stages. This is exactly the idea of ALMA-IMF, as well as some recent surveys like INFANT. \citet{cheng2024} found that a steepening trend in CMF with cloud evolution: $-0.89 \pm0.15$ for the young group and $-1.44\pm0.25$ for the evolved group. The high-mass-end steepening is also revealed in ALMA-IMF recent results (during the conference discussion; Cunningham et al. in prep.).

Evolutionary stage is nevertheless not the only driver of CMF variation. The measured slope also depends on the fitted mass range, core definition, temperature treatment, and evolutionary classification scheme. This last point is especially important for comparing large ALMA programs. As discussed during the CMF2IMF conference, evolutionary indicators are not yet homogeneous across surveys: ALMAGAL commonly uses luminosity-to-mass ratio, $L/M$, and excludes cores associated with strong radio continuum emission from CMF fitting, whereas ALMA-IMF has also used classifications based on free-free emission and compact ionized gas. These diagnostics do not necessarily select the same physical stage and may respond differently to embedded heating, feedback, multiplicity, and distance-dependent sensitivity. A meaningful comparison of CMF evolution therefore requires not only consistent fitting methods, but also a more unified calibration of evolutionary stage across surveys.

Thus, if the CMF evolves, a fitted high-mass slope is only one snapshot of a more complex and time-dependent distribution. A top-heavy, Salpeter-like, or top-light slope may reflect fragmentation physics, the evolutionary stage of the region, the prestellar-to-protostellar core fraction, and the mass interval over which the power law is fitted. Future comparisons should therefore ask not only whether a CMF is Salpeter-like, but also whether it shows a turnover, a shallow or curved component, a high-mass cutoff, or a changing prestellar-to-protostellar composition. Tracking these features with consistently defined evolutionary stages may be more informative than comparing a single fitted slope across all environments.

\section{Summary and Caveats} \label{sec:sum}

In this memo, we collect core catalogues from nearby \textit{Herschel} clouds, Galactic massive star-forming regions, CMZ clouds, and extragalactic ALMA observations, and place them into a common comparative framework. We release the Python package \texttt{CMF4All} for community use, enabling future CMF measurements to be analysed and visualized in the same context. Using this framework, we compare two choices of minimum fitting mass for the CMF power-law tail: fits starting from the observational completeness mass, $M_{\min}=M_{\rm comp}$, and fits using a uniform KS-selected threshold, $M_{\min}=M_{\rm KS}$. We find that the first choice often includes a shallower segment of the CMF, leading to apparently top-heavy slopes, whereas the second choice generally recovers slopes closer to the Salpeter value. One notable exception is ASHES, whose top-light high-mass tail may indicate that the most massive end of the core population is still being assembled at the earliest evolutionary stages.

The exercise carried out in this memo is intended as a starting point for further discussion in the community. A mass function is often treated as an answer: a slope to be measured, a reference line to be compared with, or a number to be placed beside the IMF. {\bf Yet a mass function itself is also a question.} It asks what physical processes have organized matter into the structures we observe, what history is hidden behind a present-day distribution, and what future evolution is still encoded in its shape. In this sense, the CMF should not be viewed merely as a precursor curve waiting to be shifted into the IMF, but as a fossil record of fragmentation, accretion, feedback, and environmental evolution. To understand the CMF is therefore not only to fit its slope, but to ask how dense gas is organized into stellar systems.

Several caveats remain. In addition to the fitting methodology emphasized here, CMF measurements can be biased by differences in observational setup, angular resolution, sensitivity, spatial filtering, source-extraction algorithm, temperature estimate, line-of-sight confusion, and the selection of prestellar/protostellar or bound/unbound cores. Ideally, these effects should be also homogenized, or at least explicitly quantified, when multiple surveys are compared. Due to the scope of this memo, we have not explored all of these systematics in detail. The present comparison should therefore be viewed as a first step toward a more complete framework, rather than a final homogenized measurement of the CMF across all environments.

\section{Outlook}

We propose several key points to be addressed in the near future. We acknowledge that some have already been raised during the open discussion session at the conference. 

\subsection{Improve constraints on core masses}

Millimetre continuum emission, often from a single band, is widely used to estimate core masses through
\begin{equation} \label{eq:mass_core}
M_{\rm core} = \frac{S_\nu d^2}{\kappa_\nu B_\nu(T_{\rm d})},
\end{equation}
where $S_\nu$ is the flux density, $d$ is the distance, $\kappa_\nu$ is the dust opacity, and $T_{\rm d}$ is the dust temperature. Among these quantities, $\kappa_\nu$ is the most uncertain value. But since most of ALMA surveys use the same frequency (1.3~mm), the relative uncertainty diminishes. $T_{\rm d}$ is then the second most uncertain. Some studies use gas temperatures as proxies for dust temperatures, assuming that gas and dust are thermally coupled at high densities \citep{goldsmith2001}. However, this approach becomes uncertain when no single temperature tracer is available for the full core sample, or when different tracers probe different densities, excitation conditions, and spatial scales. A systematic calibration among commonly used gas thermometers, such as CH$_3$CCH, CH$_3$CN, and NH$_3$, is therefore still needed \citep[e.g.,][]{li2026}, although whether they really trace cores is still uncertain.
An alternative approach is to constrain dust temperature directly from multi-wavelength continuum data. The ALMA-IMF team, for example, combined multiple continuum images and used the PPMAP technique to construct high-resolution temperature maps \citep[e.g.,][]{dellova2024}, although the inconsistency with gas temperature obtained from complex molecular species remains worrying \citep[e.g.,][]{motte2025}. Currently, ALMA's flexible array configurations could make it possible to obtain comparable-resolution continuum images across a wide wavelength range, from 7~mm in Band~1 to 0.35~mm in Band~10. Core-scale submillimetre/millimetre spectral energy distributions would then provide stronger constraints on both $T_{\rm d}$ and the resulting core masses.

Another caveat of single-band continuum surveys is the degeneracy between dust temperature and dust opacity. At millimetre wavelengths, cores are often assumed to be optically thin, in which case their masses can be estimated using Eq.~(\ref{eq:mass_core}). However, the optical depth of the most massive or compact cores may approach or exceed unity, leading to an underestimation of their masses if optically thin emission is assumed \citep{xu2025}. Including longer wavelengths, such as 3~mm, can help alleviate this problem because the opacity decreases approximately as $\kappa_\nu \propto \lambda^{-\beta}$, with $\beta\sim1{-}2$. 

Besides, in evolved massive star-forming regions, free-free emission from ionized gas can contaminate the millimetre continuum and artificially increase the dust flux, so multi-frequency continuum decomposition is also needed to separate dust emission from free-free contributions before deriving core masses. 

Finally, images with high dynamic range are also favoured because they can identify those faint and low-mass population objects even at the periphery of the bright sources (like massive protostellar objects). Such high dynamic range requires as complete $uv$-coverage as possible: not just using current best facility ALMA, but also including compact array like ACA 7m, and reliable single-dish data to compensate the zero-spacing missing information. The LANCET Paper~II (F. Xu et al. in prep.) combine interferometric and single-dish data; they have found that even for a small-scale dense core, without including ACA 7m data can have over 50\% flux missed in the ALMA 12m-only data. 

\subsection{Sub-fragmentation of cores}

A meaningful mapping from the CMF to the IMF requires physical constraints beyond the mass function itself. Most current ALMA CMF studies resolve cores on scales of order $2{,}000$~au, which is sufficient to identify dense fragments in massive protoclusters but still too coarse to determine how each core ultimately partitions its mass into single stars, binaries, or higher-order multiple systems \citep[e.g., see][for an example case of further sub-fragmentation of $\sim1000$~au cores down to $\sim40$~au]{palau2018}. The mapping from a core mass to a stellar mass is therefore not only an efficiency problem, but also a fragmentation and multiplicity problem. Recent work on hierarchical fragmentation in W43-MM1 and related models shows that core sub-fragmentation can reshape the mass distribution and affect the resulting stellar multiplicity, emphasizing that the IMF and multiplicity may have to be understood together rather than separately \citep[e.g.,][]{thomasson2022, motte2026}. This is also consistent with broader reviews of stellar multiplicity and clustered star formation, which highlight that multiple-system formation is a fundamental outcome of core and disk fragmentation rather than a secondary complication in particular massive star forming regions, which have higher multiplicities (see review in \citealt{offner2023} and recent surveys in \citealt{li2024} and \citealt{luo2026}). Future CMF studies therefore need to connect core-scale masses to sub-core fragmentation, protostellar multiplicity, and the final stellar-system mass function.

\subsection{Environments regulate core formation and growth}

The CMF should be studied together with the surrounding gas reservoir. A core observed at one time is not necessarily a closed mass reservoir; it may continue to accrete from filaments, clumps, and larger-scale cloud structures during protocluster formation. In this view, the core mass is a time-dependent quantity shaped by the balance between fragmentation, accretion, feedback, magnetic regulation, and dispersal. Directly measuring accretion rates onto cores remains difficult, but recent observations increasingly show that streamer-like or filamentary flows can connect cloud, core, and disk scales. Examples include chemically fresh streamers feeding young protostellar systems \citep{pineda2020}, multi-scale accretion flows in massive hub-filament systems \citep{xu2023}, and ``V-shaped'' gas kinematics reported by the ALMA-IMF team \citep[e.g.,][and ongoing work by Amelia Stutz's team]{alvarez2024, sandoval2025, salinas2025}. Detailed pixel-wise multi-component fitting is still largely limited to case studies, but statistical approaches are beginning to connect these flows to core properties. For example, ALMAGAL measurements of filamentary flow rates toward cores show how accretion can be analysed as a function of evolutionary stage, distance from the core, and core mass \citep{wells2024}. In addition, core-scale infall rates can be inferred statistically from asymmetric line profiles \citep[e.g.,][]{morii2026b}.

This environmental view also motivates the need for large mosaic observations. A single ALMA pointing centred on the brightest continuum peak can provide a detailed view of the densest region, but it may miss the broader population of cores forming along filaments, ridges, and lower-density cloud structures. If core formation is fed by large-scale gas flows, then the relevant physical unit is not only the compact protocluster centre, but the entire connected gas reservoir from which cores form and grow. Large mosaics are therefore essential for measuring a more complete core population, sampling both central and distributed modes of core formation, and determining whether the CMF varies with position relative to hubs, filaments, feedback fronts, or gravitational potential minima. They also reduce the risk that a CMF is biased toward the most active or luminous part of a cloud. In this sense, a complete CMF requires not only sufficient sensitivity and angular resolution, but also sufficient spatial coverage to capture the full core-forming environment. The ALMA-IMF, LANCET surveys, ALMA PI projects (2025.1.00044.S and 2023.1.01082.S), and ALMA large program Panta-Rei (2025.1.00383.L) are providing promising data sets for this scientific goal.

Environmental magnetic fields provide another important link between cores and their surroundings \citep[see review in][]{hull&zhang2019}. They can regulate core formation and growth by dissipating turbulence, stabilizing filaments, and delaying collapse. These effects may operate differently across a protocluster, producing age spreads or accretion timescales among cores formed within the same parent cloud \citep{xu2026b}. Therefore, the CMF should not be treated as a purely local outcome of density fragmentation, but as the result of continuous interaction between dense cores and their environments. Studies that combine core masses, accretion flows, magnetic-field structure, evolutionary stage, and large-scale spatial coverage are essential for turning the CMF from a static distribution into an evolutionary diagnostic.

\subsection{Direct stellar mass spectra from ionizing photons}

A complementary way to connect the CMF to the IMF is to move beyond dense cores and directly census the newly formed massive stars. Deep centimetre continuum observations can detect free-free emission from compact, ultracompact, and hypercompact H{\sc ii} regions, which are powered by massive stars that have reached, or are close to, the zero-age main sequence. 
Assuming that most of the centimetre compact sources in massive star-forming regions are associated with ultra/hyper-compact HII regions, the radio luminosity function could serve as a first proxy to the IMF in the high-mass regime. The preliminary work by Yanza (2025\footnote{see Sec. 5 of \url{https://tesiunamdocumentos.dgb.unam.mx/ptd2025/abr_jun/0871918/Index.html}.}) explored the radio luminosity function of several massive star-forming regions and found hints of a potential steepening of the slope with time. This could suggest that further low-mass star formation is taking place after the formation of massive stars, a trend worth exploring in more detail. However, the radio luminosity function does not involve a direct measurement of the stellar mass. To convert the centimetre flux to a measurement of stellar mass,
%
the observed radio continuum can be used to derive the emission measure and the Lyman-continuum photon rate, $Q_0$, which can then be converted into spectral type and stellar mass using stellar-atmosphere or population-synthesis calibrations \citep{Ho&Haschick1981}. 

A recent VLA study of W49A by Roberto Galv{\'a}n-Madrid, Mariana Ju{\'a}rez-Gama, and collaborators demonstrates the potential of this method. Using 3.3~cm data at a physical resolution of $\sim2{,}000$~au, they identified 101 robust compact ionized sources and inferred the embedded zero-age massive stellar population from their ionizing-photon rates (Ju{\'a}rez-Gama et al., under review). Their work highlights both the promise and the caveats of radio-selected stellar mass functions. On the one hand, ionizing photons provide a more direct probe of the newly formed massive stellar population than dust cores. On the other hand, the observed radio-selected mass function may be affected by optical depth, dust absorption of ionizing photons, unresolved multiplicity, mass-dependent \ion{H}{ii}-region lifetimes, and ongoing accretion. In particular, very massive stars may disperse their compact \ion{H}{ii} regions rapidly, while rapidly accreting massive protostars may remain bloated and too cool to produce strong ionizing flux \citep[e.g.,][]{pandey2025}. After correcting for optical-depth effects, their results suggest a top-light high-mass stellar mass function over the stellar-mass range of $\sim10$--$30~\msun$, implying a relative deficit of the most massive radio-detected \ion{H}{ii} regions. This behaviour is qualitatively consistent with the possibility illustrated in Fig.~\ref{fig:cmf_segmentation}, in which the highest-mass end of the mass function may deviate from a smooth power law and instead approach an exponential-like cutoff. With current facilities such as the JVLA and ATCA, and in the near future with SKA-Mid, Galactic-plane surveys reaching frequencies of $\sim10$--$20$~GHz will provide a powerful way to extend this approach to statistical samples \citep[e.g.,][]{skagp}.

\subsection{From ISM density structure to observed CMFs}

If a core is not a fundamental unit of nature, but a practical way of naming structure in a continuous medium, then the CMF is not only a physical distribution but also an observational construction. A recurring question raised during the conference is whether a ``core'' has a unique physical definition, or whether it is partly an observer-defined density enhancement within a turbulent and hierarchical interstellar medium. If the latter is true, then the CMF should not be interpreted independently of the underlying gas-density field from which cores are extracted. This perspective is reminiscent of the column-density probability distribution function, or N-PDF, whose high-column-density power-law tail is known to develop and evolve as gravity becomes increasingly important \citep[e.g.,][]{schneider2022, jiao2025}. It is therefore natural to ask whether the evolution of the CMF is, at least partly, an observationally discretized manifestation of the evolving N-PDF: as the gas density field develops a stronger high-density tail, source-extraction algorithms identify more dense peaks, and the resulting core mass distribution changes accordingly.

This question can be addressed most cleanly through a simulation-to-observation workflow, which is also the central idea of the {\it Rosetta Stone} project \citep{rosetta1, rosetta2, rosetta3}. Starting from numerical simulations of cloud and protocluster formation, one can construct synthetic dust-continuum images through radiative-transfer post-processing, including realistic temperature gradients, optical-depth effects, and viewing-angle dependence. These intrinsic images can then be passed through ALMA observation simulators to account for angular resolution, sensitivity, interferometric filtering, noise, and $uv$ coverage. The resulting synthetic observations should finally be analysed with the same core-extraction, classification, completeness estimation, and CMF-fitting procedures used for real data. Unlike observations, however, synthetic observations provide access to the full evolutionary history of the individual parent clumps, allowing the intrinsic evolution of the CMF to be followed without relying on observational evolutionary indicators such as the L/M parameter (although it has proved to be a reliable evolutionary indicator; \citealt{rosetta2}) or alternative classification schemes. Furthermore, simulations make it possible to systematically investigate how different initial conditions of molecular-cloud collapse, such as e.g., level of turbulence or magnetic-field strength, affect the resulting CMF and its evolution.

Such a workflow also provides a way to test what an observed core actually corresponds to in three dimensions. A consistent comparison should be made between cores identified in projected 2D continuum images, density structures identified directly in 3D simulation cubes, and physically motivated structures selected by energy-based or virial-theorem criteria \citep[see recent \texttt{vibes} code in][]{chevalier2026}. This comparison is essential because a projected emission peak is not necessarily the same as a gravitationally bound entity, and a bound 3D structure may not always appear as an isolated core in an interferometric image. Ideally, the term ``core'' should refer not only to a detectable density enhancement, but to a physically meaningful unit with some likelihood of forming one or more stars. Forward-modelling approach would allow us to separate physical evolution from observational definition. It can test whether a measured CMF slope reflects the intrinsic fragmentation hierarchy of the gas, the time evolution of the density field, the dynamical boundedness of structures, or the way cores are identified in projected continuum images. 

On the theoretical side, to reproduce full mass spectrum of cores are encouraged. For example, \citet{liu2025b} suggest that CMF can be driven by filament fragmentation. In their theory, filaments and fibres are the building blocks of ISM, and they develop a fractal and turbulent tree with a fractal dimension of 2 and a Larson's law exponent $\beta$ of 0.5. The fragmentation driven by convergent flows along the splines of the fractal tree successfully reproduce a Kroupa-IMF-like CMF. \citet{gjergo2026} apply the Maximum Entropy principle to the fragmentation of star-forming clumps and recover the power-law form of the IMF. Their result shows that any distribution deviating from the IMF can violate the Maximum Entropy principle and inspires a first-principles foundation for the deterministic nature of star formation.

\section*{Postlude}

The more we have learned, the more we realize how much remains unknown. Nearly three decades after the first discoveries that the CMF resembles the IMF, the connection between dense cores and stars remains unresolved, and perhaps even less straightforward than it once seemed. The origin of the stellar IMF is not written in a single slope, nor hidden in a simple shift of the CMF by a constant efficiency factor. It is written in the full history of how molecular gas fragments, accretes, collapses, forms multiple systems, and responds to feedback, magnetic fields, and its larger-scale environment. The CMF is therefore not merely the shadow of the IMF cast backward in time. It is a present-day imprint of a cloud still evolving, a record of both what has already formed and what has not yet become stars.

This also means that the CMF should not be reduced to one fitted number. Its turnover, shallow components, Salpeter-like regime, and possible high-mass cutoff may each carry different physical and observational meanings. At the same time, even the word ``core'' may not refer to a uniquely defined physical unit, but to a structure extracted from a continuous, turbulent, and evolving interstellar medium. To understand the CMF, we therefore need not only deeper and sharper observations, but also better core-mass estimates, better links between cores and their environments, direct probes of the emerging stellar population, and simulation-to-observation workflows that reproduce how CMFs are actually measured. We hope this memo can encourage future instrumental upgrades, observational campaigns, and numerical experiments to turn the CMF from a static distribution into a physical and evolutionary diagnostic of how dense gas becomes stellar systems.

\backmatter

\bmhead{Data availability}

All the core catalogues excepted for unpublished QUARKS survey used in this work are publicly available in \url{https://github.com/XFengwei/CMF4All/tree/main/cmf_zoo/data}. The LANCET core catalogue will be soon published in the LANCET paper series, but we also put it into the release here. Please contact us for scientific usage.

\bmhead{Code availability}

All the code used for calculation and plotting is compiled as a package called \texttt{CMF4All}, which is publicly available in \url{https://github.com/XFengwei/CMF4All}.

\bmhead{Acknowledgements}

We acknowledge Wenyu Jiao and Zhengying Zhang for their providing the core catalogues. We sincerely appreciate the interesting discussions with Fr{\'e}d{\'e}rique Motte, Ashley Barnes, Adam Ginsburg, Amelia Stutz, Katharina Immer, Philippe Andr{\'e}, Enrique Vazquez-Semadeni, and Paolo Padoan during the conference. We appreciate the SOC of the conference, especially Morten Andersen and Thomas Stanke.

\bibliography{ref}

\end{document}